\documentclass[twoside,twocolumn,9pt]{article}
\usepackage{extsizes}
\usepackage[super,sort&compress,comma]{natbib} 
\usepackage[version=3]{mhchem}
\usepackage[left=1.5cm, right=1.5cm, top=1.785cm, bottom=2.0cm]{geometry}
\usepackage{balance}
\usepackage{times,mathptmx}
\usepackage{sectsty}
\usepackage{graphicx} 
\usepackage{lastpage}
\usepackage[format=plain,justification=justified,singlelinecheck=false,font={stretch=1.125,small,sf},labelfont=bf,labelsep=space]{caption}
\usepackage{float}
\usepackage{fancyhdr}
\usepackage{fnpos}
\usepackage[english]{babel}
\addto{\captionsenglish}{%
  
}
\usepackage{array}
\usepackage{droidsans}
\usepackage{charter}
\usepackage[T1]{fontenc}
\usepackage[usenames,dvipsnames]{xcolor}
\usepackage{setspace}
\usepackage[compact]{titlesec}
\usepackage{hyperref}

\usepackage{epstopdf}
\usepackage{mathrsfs,amsmath}

\definecolor{cream}{RGB}{222,217,201}

\usepackage{amssymb}

\newcommand{\rhox}{\rho(\hat{x})}
\newcommand{\rhob}{\bar{\rho}}
\newcommand{\etax}{\eta(\hat{x})}
\newcommand{\etab}{\bar{\eta}}
\newcommand{\F}{\mathscr{F}}
\newcommand{\x}{\hat{x}}
\newcommand{\dx}{\mathrm{d}\x}
\newcommand{\Y}{\mathrm{Y}}
\newcommand{\hcoeff}{\sqrt{\frac{4\pi}{2\ell+1}}}

\begin{document}

\pagestyle{fancy}
\thispagestyle{plain}
\fancypagestyle{plain}{
}

\makeFNbottom
\makeatletter
\renewcommand\LARGE{\@setfontsize\LARGE{15pt}{17}}
\renewcommand\Large{\@setfontsize\Large{12pt}{14}}
\renewcommand\large{\@setfontsize\large{10pt}{12}}
\renewcommand\footnotesize{\@setfontsize\footnotesize{7pt}{10}}
\makeatother

\renewcommand{\thefootnote}{\fnsymbol{footnote}}
\renewcommand\footnoterule{\vspace*{1pt}%
\color{cream}\hrule width 3.5in height 0.4pt \color{black}\vspace*{5pt}} 
\setcounter{secnumdepth}{5}

\makeatletter 
\renewcommand\@biblabel[1]{#1}            
\renewcommand\@makefntext[1]%
{\noindent\makebox[0pt][r]{\@thefnmark\,}#1}
\makeatother 
\renewcommand{\figurename}{\small{Fig.}~}
\sectionfont{\sffamily\Large}
\subsectionfont{\normalsize}
\subsubsectionfont{\bf}
\setstretch{1.125} 
\setlength{\skip\footins}{0.8cm}
\setlength{\footnotesep}{0.25cm}
\setlength{\jot}{10pt}
\titlespacing*{\section}{0pt}{4pt}{4pt}
\titlespacing*{\subsection}{0pt}{15pt}{1pt}

\fancyfoot{}
\fancyfoot[RO]{\footnotesize{\sffamily{1--\pageref{LastPage} ~\textbar  \hspace{2pt}\thepage}}}
\fancyfoot[LE]{\footnotesize{\sffamily{\thepage~\textbar\hspace{3.45cm} 1--\pageref{LastPage}}}}
\fancyhead{}
\renewcommand{\headrulewidth}{0pt} 
\renewcommand{\footrulewidth}{0pt}
\setlength{\arrayrulewidth}{1pt}
\setlength{\columnsep}{6.5mm}
\setlength\bibsep{1pt}

\makeatletter 
\newlength{\figrulesep} 
\setlength{\figrulesep}{0.5\textfloatsep} 

\newcommand{\topfigrule}{\vspace*{-1pt}%
\noindent{\color{cream}\rule[-\figrulesep]{\columnwidth}{1.5pt}} }

\newcommand{\botfigrule}{\vspace*{-2pt}%
\noindent{\color{cream}\rule[\figrulesep]{\columnwidth}{1.5pt}} }

\newcommand{\dblfigrule}{\vspace*{-1pt}%
\noindent{\color{cream}\rule[-\figrulesep]{\textwidth}{1.5pt}} }

\makeatother

\twocolumn[
  \begin{@twocolumnfalse}
\vspace{3cm}
\sffamily
\begin{tabular}{m{4.5cm} p{13.5cm} }

& \noindent\LARGE{\textbf{Phase diagram of SALR fluids on spherical surfaces}} \\
\vspace{0.3cm} & \vspace{0.3cm} \\

 & \noindent\large{Stefano Franzini,\textit{$^{a}$} Luciano Reatto\textit{$^{a}$} and Davide Pini\textit{$^{a}$}} \\

& \noindent\normalsize{ We investigate the phase diagram of a fluid of hard-core disks 
confined to the surface of a sphere and whose interaction potential contains a short-range attraction followed by a long-range repulsive tail (SALR). Based on previous works in the bulk we derive a stability criterion for the homogeneous phase 
of the fluid, and locate a region of instability linked to the presence of a negative minimum in the spherical harmonics
expansion of the interaction potential.
The inhomogeneous phases contained within this region are characterized using a mean-field density functional theory. We
find several inhomogeneous patterns that can be separated into three broad classes: cluster crystals, stripes, and
bubble crystals, each containing topological defects. Interestingly, while the periodicity of inhomogeneous phases at
large densities is mainly determined by the position of the negative minimum of the potential expansion, the finite  
size of the system induces a richer behavior at smaller densities. } \\

\end{tabular}

 \end{@twocolumnfalse} \vspace{0.6cm}

  ]

\renewcommand*\rmdefault{bch}\normalfont\upshape
\rmfamily
\section*{}
\vspace{-1cm}


\footnotetext{\textit{$^{a}$~Dipartimento di Fisica ``A.~Pontremoli'', Universit\`a di Milano, Via Celoria 16, 20133 Milano, Italy. E-mail: stefano.franzini@sissa.it}}





\section{INTRODUCTION}

Colloids interacting through a combination of a short-range attraction and a long-range 
repulsive tail (SALR potential) can be obtained by combining electrostatic interactions 
described by the DLVO theory, which provide the long-range repulsive part, and depletion 
effects due to non-adsorbing crowders, such as polymers interspersed in the solution, 
which create a tunable short-range attraction \cite{Sedgwick_2004}\cite{Mani_2014}. 
The interest in fluids with competing interactions stems from the theoretical prediction 
that their phase diagram would contain a plethora of inhomogeneous mesophases, such as 
cluster or bubble crystals, but also cylinders, lamellae, and even bicontinuous structures 
such as gyroids \cite{Pini_2017, Ciach_2008, Ciach_2010, Ciach_2013, Edelmann_2016}, 
although sampling some of these phases experimentally is still an open challenge \cite{Royall_2018}. 

The richness of this phase diagram is possible thanks to the unique competition between 
the attractive part of the potential, which induces the aggregation of the colloids, and 
its repulsive tail, which prevents global phase separation and promotes the formation of 
aggregates of limited size instead.

This phenomenon can also be observed in two-dimensional fluids of colloids adsorbed on a 
surface, for example on a liquid-liquid or air-liquid interface, in which case the phase 
diagram contains clusters, stripes and bubbles \cite{Archer_2008}. 
Here attractive interactions can be mediated by the substrate though a variety of 
mechanisms such as capillary interactions, lateral depletion forces or Casimir 
attractions \cite{Dan_1993}\cite{Netz_1997}\cite{van_der_Wel_2016}\cite{Destainville_2018}, 
while the repulsive part of the potential can still be explained by electrostatic forces.

Of particular interest to the present study are models in which particles of the SALR 
fluid are constrained to move on a curved surface, such as that of a sphere. These 
models can be used to represent protein inclusions on the surface of biological 
vesicles or micelles, or pickering emulsions, where solid colloids adsorb at the 
interface between two liquid phases (water and oil for example) \cite{Rozynek_2014}\cite{Thompson_2015}. 
Previous studies have shown that SALR fluids on a spherical surface still form patterns 
similar to the ones found in the planar case, providing a mechanism to obtain vesicles 
or colloidosomes covered with anisotropic patches with a controllable geometry \cite{Zarragoicoechea_2009}\cite{Meyra_2010}.

The possibility of exploiting the self-assembling properties of these fluids on a spherical 
substrate as a path to bottom-up production of patchy particles motivates us to study their 
phase diagram. Patchy particles are colloids that display anisotropic interactions either 
due to their shape or because of the presence on their surface of functionalized patches, 
from which they take their name \cite{Pawar_2010}. They find applications in multiple fields, 
such as in medicine, where they get used for targeted drug delivery \cite{Poon_2010}, in 
material science, as building blocks for hierarchically self-assembled structures \cite{Fang_2009}\cite{C_G_rlea_2017}, 
and in industry, where they are used in a variety of emerging technologies \cite{Liddell_2003}. 
One of the major challenges in the field is the scalability of production techniques. 
Top-down manufacturing strategies, such as photolithography, tend to incur into size and 
processing limitations \cite{Pawar_2010} that new bottom-up techniques seek to overcome by 
self-assembling the target anisotropic particle \cite{Glotzer_2004}. 

Previous studies on SALR fluids on a sphere \cite{Zarragoicoechea_2009}\cite{Meyra_2010} focused 
on describing the patterns forming at different temperatures and densities using Monte Carlo 
simulations, however no phase diagrams have been proposed. Hence we use a mean field density 
functional theory to obtain the phase diagram of the SALR fluids. We show that at large enough
densities the patterns can be predicted  from the harmonic order of the absolute negative
minimum found in the spherical harmonics series of the non singular part of the potential.

The outline of the paper is as follows: in section 2 we will present the details of the SALR model,
as well as the DFT-functional and Monte Carlo simulations we have used to study it; in section 3 we
will first review results about the homogeneous phase and its instability region, we will then present a
thorough investigation of the phase diagram according to DFT, and finally we will show some of the
metastable phases and discuss the difference with respect to soft-core fluids forming microphases (such as the
generalized gaussian model of exponent 4, or GEM4 \cite{Mladek_2006}); in section 4 we will summarize our findings and discuss future perspectives.

\section{METHODS}

\subsection{The SALR model}

In our model we describe particles constrained to a spherical surface and interacting via 
a two-body, isotropic potential $v(r)$ given by a hard core followed by a short-range 
attractive and long-range repulsive (SALR) tail $w(r)$, with $r$ being the euclidean distance, 
which we model as the sum of two Yukawa functions of opposite signs:

\begin{equation}
        v(r)=
    \begin{cases}
    \displaystyle{\infty}, & \text{if}\ r<\sigma^* \\
      -\displaystyle{\frac{R \varepsilon}{r}}\bigg[ \exp\big(-\gamma_1 r/\sigma^* \big) - A\exp\big(-\gamma_2 r/\sigma^* \big) \bigg], & \text{otherwise}
    \end{cases}
    \label{pot}
\end{equation}

This definition introduces four different lengthscales: $R$ is the radius of the 
sphere; the diameter $\sigma^* = R\sqrt{2-2\cos\frac{\sigma}{R}}$ is defined in terms of the
curvilinear diameter $\sigma$, i.e. the minimum geodesic distance two particles can achieve
without overlapping; $\gamma_1/\sigma^*$ and $\gamma_2/\sigma^*$ are the inverse of the 
attractive and repulsive ranges respectively. We will measure lengths in units of $R$ and  
set $\gamma_1=\frac{5}{6}$ and $\gamma_2=\frac{1}{2}\gamma_1$, so that we only 
need to specify one lengthscale, namely $\sigma$.

The other quantities which define the potential are the attractive amplitude $\varepsilon$, 
which we set to $1$, and the repulsive amplitude $A$. Here we follow 
a previous work \cite{Pini_2017}: we choose $A$ so that $\int\dx\, w(\x) = 0$. In order 
to do so, we adopt the standard prescription of fixing the tail inside the hard-core 
region to its minimum, $w(\sigma^*)$. This gives us the following expression for the amplitude:

\begin{equation}
    A = \frac{1}{2}\frac{\big(\gamma_1 + 2 \big) e^{-\gamma_1} - 2 e^{-2 R\gamma_2/\sigma^*}}
    {\big( \gamma_2 + 2 \big) e^{-\gamma_2} - 2 e^{-2 R \gamma_1/\sigma^*}},
\end{equation}

which gives $A \simeq 0.3866$ at $\sigma/R = 0.1$ for the values of $\gamma_1$ and $\gamma_2$ used here.

\begin{figure}[t]
\centering
  \includegraphics[width=0.48\textwidth]{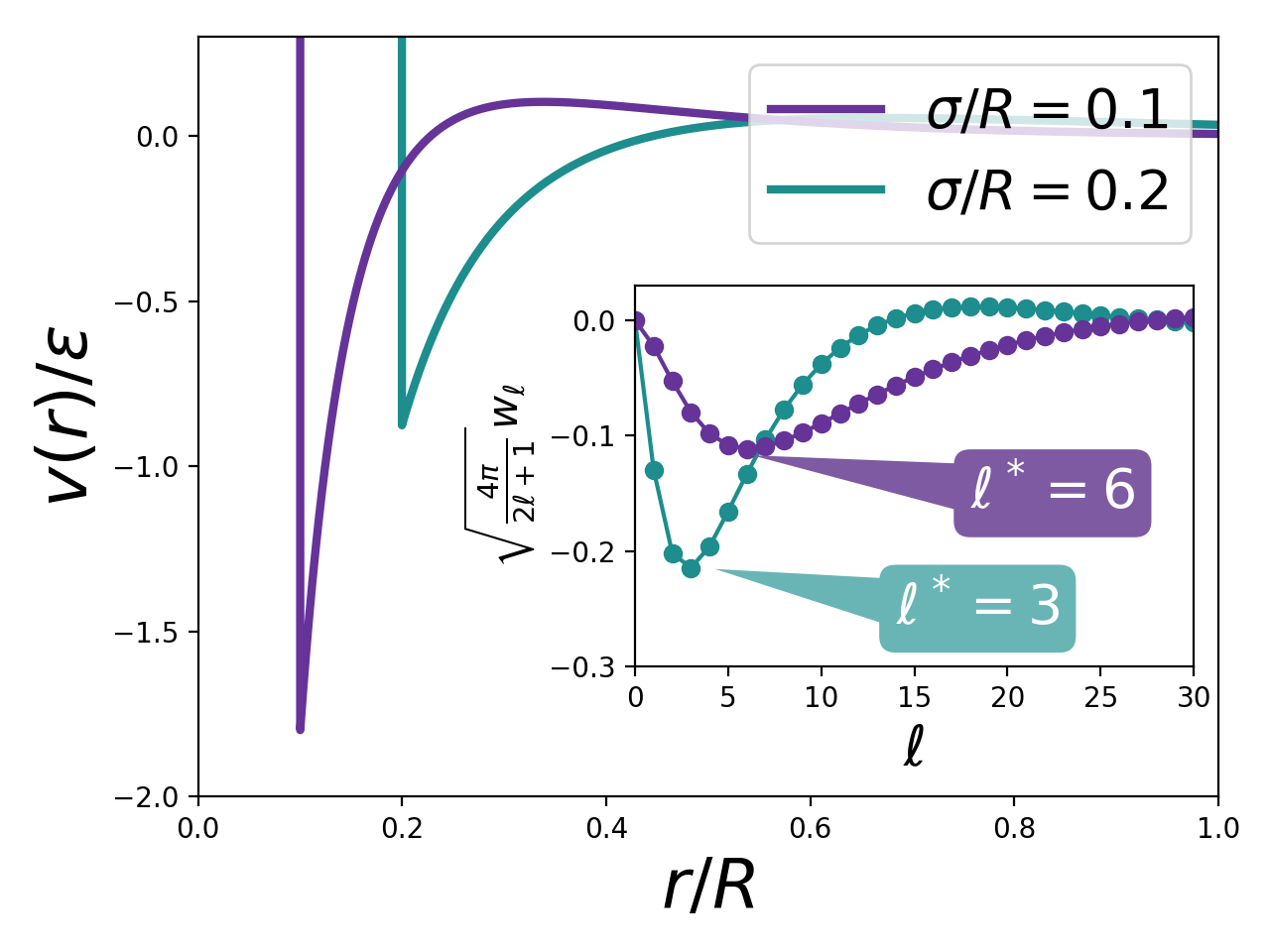}
  \caption{The two-body interaction potential at $\sigma/R = 0.1$ and $\sigma/R = 0.2$. The inset shows the spherical harmonics coefficient of the SALR tail: because the potential is isotropic the spherical harmonics expansion is reduced to its zonal components $w_{\ell} = \int \dx w(\x)\Y_{\ell,0}(\x)$, where $\Y_{\ell,0}(\x)$ are the zonal spherical harmonics. The minima in the two potentials are indicated in the balloons beside the plots: they correspond to $\ell^*=6$ at $\sigma/R=0.1$, and $\ell^*=3$ at $\sigma/R=0.2$. }
  \label{competing_pot}
\end{figure}

Examples of the potential can be seen in figure \ref{competing_pot}, where we also plot 
the spherical harmonics coefficients of the SALR tail, defined by:

\begin{equation}
    w_{\ell} = \int \dx\, w(\x)\Y_{\ell,0}(\x),
\end{equation}
where $\x$ represents the coordinates of a point on the surface of a unitary sphere. Notice we only need
to retrieve the $m=0$ components using the zonal spherical harmonics $Y_{\ell,0}(\x)$,
thanks to the rotational symmetry of the potential. Moreover we consider real spherical harmonics,
since all the functions we consider are real. These coefficients display negative minima at $\ell^*=6$ 
and $\ell^*=3$ respectively for $\sigma/R=0.1$ and $\sigma/R=0.2$: we will show in the next section that 
this property leads to the formation of microphases.

We also introduce here the other quantities which we are going to use throughout the paper. 
The density of $N$ particles on the sphere is given by $\rho = N/(4\pi R^2)$. For simplicity 
we shall use the adimensional quantity $\rho R^2$. Since the particles have a finite size 
$\sigma$ we also introduce the packing fraction $\eta$ which is the ratio between the total 
surface area of the sphere and the area occupied by the particles. Using the equation for 
the area of a spherical cap of curvilinear diameter $\sigma$ we obtain $\eta = 2\pi R^2 ( 1 - \cos \frac{\sigma}{2 R} )\rho$. 
Notice that while in principle $\eta$ can vary between $0$ and $1$, the actual upper 
limit is given by the close packing fraction of the disks, which on a plane would be 
achieved by the hexagonal lattice packing $\eta_{CP} = \frac{\pi\sqrt{3}}{6} \sim 0.9069$. 
The problem of finding the optimal packing of $N$ disks on a spherical surface is still 
open, but since we will consider small $\sigma/R$ ratios we can consider the planar 
close packing as an appropriate approximate upper boundary for the packing fraction $\eta$. 
We measure the temperature of the system in reduced units $T^* = k_B T/ \epsilon$, where $k_B$ is the Boltzmann constant. Notice that some
of the attractive effective interactions that may lead to the SALR potential can be described by athermal
phenomena (such as depletion), so the inverse temperature $\beta=\epsilon/k_B T$ can also be interpreted as a concentration
of depletants. Finally, since we will work in the 
grand canonical ensemble, we also introduce the chemical potential $\mu$.

\subsection{Density functional theory}
\label{sec:theory}

The SALR potential is well known to form clusters and other inhomogeneous phases \cite{Pini_2017}\cite{Archer_2008}. 
The attractive part of the potential helps particles aggregate when the density of the fluid is large enough, 
however the repulsive tail keeps the size of the agglomerates finite, leading to a complex behavior 
in which particles arrange in a variety of different patterns depending on the thermodynamic state of the system.

In the next section we will build the tools needed to make this qualitative argument for 
the emergence of inhomogeneous phases in SALR fluids more quantitative and obtain a 
phase diagram of the system, similarly to what was done for soft-core fluids on a sphere \cite{Franzini_2018}.

To describe this system we use a density functional theory (DFT), which expresses 
the grand potential $\beta\Omega$ of the fluid as a functional of its density 
profile $\rhox$. For interacting systems the exact functional is generally unknown, 
so we resort to an approximation where a simple mean field perturbation is applied 
to a reference functional which contains the ideal gas and the hard-core contributions. 
The former is known exactly, but the latter must again be obtained from some sort 
of approximation. A local density approximation (LDA) for the free energy of a fluid 
of hard disks can be obtained from equation of states known in the literature, which 
in turn have been derived by a variety of tools such as virial expansion and integral 
equations. Here we follow Archer \cite{Archer_2008} in using the scaled-particle approximation. 
This gives us the following grand potential functional defined on the unit sphere $S^2$

\begin{equation}\label{functional}
\begin{aligned}
    \beta\Omega[\rho] = \beta\F_{id} + \beta\F_{HD} + \beta\F_{P} + \beta\F_{ext} - \beta\mu N \\
    \beta\F_{id} -\beta\mu N = \int_{S^2} \dx\, \rhox \big[ \ln( \rhox \delta_{TH}^2 ) - 1 - \beta\mu\big]\\
    \beta\F_{HD} = \int_{S^2} \dx\, \rhox \big[ (1 - \etax)^{-1} - \ln( 1 - \etax) -1 \big]\\
    \beta\F_{P} = \frac{1}{2} \beta \int_{S^2} \int_{S^2} \dx \dx'\, \rhox \rho(\x') w(\x,\x')\\
    \beta\F_{ex} = \beta \int_{S^2} \dx\, \rhox \phi(\x)
\end{aligned}
\end{equation}
where $\F_{id}$ is the ideal-gas free energy, $\F_{HD}$ is the hard-disk 
contribution, $\F_{ext}$ is the contribution due to the action of an external 
field $\phi$ and $\F_{P}$ is the excess free energy due to the contribution of the interaction potential. We have also introduced 
the thermal length $\delta_{TH} = \sqrt{\frac{h^2 \beta}{2\pi m}}$, $h$ 
being the Planck constant and $m$ being the mass of a particle.

It should be noted that the functional we employ is quite crude with respect to more 
accurate functionals available in the literature, both for the hard-disk contribution
and for the excess free energy term. For example, fundamental measure theory provides reference 
functionals for the hard-disk contributions that can accurately probe the fluctuations 
of the density profile even on the lengthscale of the disk diameter \cite{Roth_2012}, while the LDA 
functional used here will only provide useful information about the profile on the much 
larger lengthscale of the meso-structures found in the system \cite{Archer_2008}. Moreover we derive the 
reference functional from an equation of state which is valid for a planar system, and 
one may be concerned that a spherical system would require corrections to this equation 
of state and hence to the functional: however numerical solutions of the Percus-Yevic equation for
hard-disks on a sphere surface \cite{Lishchuk_2006} show that the correction
is of second order on $\sigma/R$ and does not have a large impact for the relatively small values 
of this parameter considered here. 

On the other hand, the excess free energy due to the 
non singular part of the potential which we use is known to be inaccurate when describing
the critical point of the phase diagram, because it is the result of a mean field approximation. 
This inaccuracy was brought forth in a simulation study of a three-dimensional SALR fluid in the bulk \cite{Charbonneau_2016}, which showed that the phase portrait is qualitatively different from the mean-field picture, especially as the high-temperature limit of the inhomogeneous phases is reached, and thermal fluctuations become important. In fact, both approaches predict the same kind of periodic structures, namely, cluster, tubular and lamellar phases in order of increasing density. 
However, according to the mean-field DFT all of the above phases survive up to the highest temperature at which the system is inhomogeneous, whereas in the simulation, as the temperature increases, the cluster phase melts first, to be followed by the tubular phase, and finally by the lamellae. Moreover, the whole inhomogeneous domain is shifted to higher densities compared to the mean-field DFT predictions.

The fact that these discrepancies should be traced back to the mean-field approximation was made clear in a subsequent theoretical investigation \cite{Ciach_2018}, in which fluctuations are taken into account on the top of the mean-field result, and the phase diagram thus obtained displays the same qualitative features obtained in the simulations.

All the above concerns are well funded, but here we opt for simplicity, even though this may affect
the accuracy of our findings. We will not be able to describe the internal structure 
of clusters and other meso-structures, and we will not obtain an accurate description of the 
critical point in the phase diagram using this functional. Nevertheless we will be able to 
compute a qualitative phase diagram of the system. We will also be able to compare our results
for the homogeneous fluid to simulations we performed.

Once we have a functional, we can obtain the equilibrium density profile by solving the 
Euler-Lagrange equations in search of a global minimum of the thermodynamic potential:

\begin{equation}\label{EulerLagrange}
\begin{aligned}
    \frac{\delta \beta\Omega[\rho]}{\delta \rhox} = 0.
\end{aligned}
\end{equation}

In general these equations do not have a single solution for a given thermodynamic state, 
meaning that the system has different metastable states, so one needs to obtain the profiles 
of different minima and compare their grand potentials. Moreover these equations are not 
usually analytically solvable, so the results are obtained through numerical methods \cite{Pini_2015}\cite{Pini_2017}\cite{Franzini_2018}. 

The homogeneous fluid is a special case in which one can obtain an analytical solution by 
plugging in the equations a flat density profile. Using a homogeneous density does not 
always lead to a stable solution, that is, the homogeneous fluid may not be a local 
minimum of the grand potential. However one can always recast the functional in terms of 
the density of the homogeneous solution $\rhob$ and its packing fraction $\etab$, instead of the chemical potential $\mu$. 
From the homogeneous solution of the Euler-Lagrange equations we obtain

\begin{equation}
\begin{aligned}
\beta \mu = \ln( \rhob \delta_{TH}^2 ) + \beta\mu^{ex},\\
\beta \mu^{ex} \equiv \frac{3\etab - 2\etab^2}{(1 - \etab)^2} - \ln(1-\etab) + \int_{S^2} \dx\,w(\x),
\end{aligned}
\end{equation}
where we have introduced the excess chemical potential $\mu^{ex}$, which allows us to rewrite the ideal gas reference functional as 

\begin{equation}
    \beta\F_{id} - \beta\mu N = \int_{S^2} \dx\, \rhox \big[ \ln( \rhox/\rhob ) -1 -\beta\mu^{ex} \big]. 
\end{equation}

Since we will study the phase diagram in terms of the density $\rho$ of the fluid, 
rather than its chemical potential $\mu$, this parametrization of $\mu^{ex}$ will
simplify the task of exploring the thermodynamic states of the system, by allowing
us to obtain an approximate location of each sampled thermodynamic state on the phase diagram.

In order to solve the Euler-Lagrange equations we adapt the minimization algorithm 
previously developed for bulk \cite{Pini_2017} and spherical systems \cite{Franzini_2018} 
of soft particles, which uses the SHTOOLS python module to compute spherical harmonics \cite{SHTOOLS, Frigo_2005}
. This algorithm discretizes the density profile over a $K \times 2K$ grid 
of equispaced points in $\theta$ and $\phi$, 
and proceeds to minimize the grand potential of the discretized 
density profile using an optimized version of the gradient descent algorithm. In 
this way we allow the algorithm to freely explore virtually any density profile, instead 
of limiting ourselves to a set of profiles obtained \textit{a priori}. This is especially 
useful because the sphere topology prevents the formation of perfect lattices and the 
positions of the disclinations defects can be difficult to guess \textit{a priori}. Moreover, 
it lets us explore unexpected metastable patterns, such as spiral phases, which would 
be discarded otherwise. The trade off is that the algorithm still needs a set of initial 
density profiles from which to start the gradient descent, which must be chosen so as to 
explore as much of the free energy landscape as possible. We employ different starting 
conditions using (i) random patterns, (ii) striped patterns, (iii) high density spots at 
the vertices of regular polyhedra, (iv) low density spots at the vertices of regular polyhedra, 
(v) profiles obtained from previous runs of the minimization algorithm. Even with this 
setup, it must be noted that there is no absolute guarantee that the final density profile 
is truly the global minimum. However by comparing our results with previous ones we can 
achieve a good degree of confidence in their overall correctness.
We set the parameter $K$ determining the number of grid points at $K=256$. For this choice of $K$, the algorithm is still rapidly convergent, as shown in figure S3 of the Supporting Material, while at the same time its results are basically unaffected by the discretization. In fact, halving or doubling $K$ does not lead to appreciable changes in the grand potential at convergence, as displayed in figure S4 of the Supporting Material. Moreover, we show in figures S5, S6, and S7 of the Supporting Material that
the equilibrium distribution at three different densities is also unaffected by halving or doubling $K$.

We still need to show that the system allows the formation of equilibrium inhomogeneous 
patterns. To do so, consider the functional we have described: the trivial homogeneous 
solution to the Euler-Lagrange equations is not necessarily a minimum of the grand potential. 
For this to be the case, it must satisfy the additional condition that the density profile 
$\rhob$ is stable with respect to arbitrary small fluctuations $\delta\rhox$

\begin{equation}\label{stability}
    \int_{S^2} \int_{S^2} \dx\dx'\,\delta\rhox\delta\rho(\x')c(\x,\x') < 0 \, ,
\end{equation}
where we introduced the direct correlation function $c(\x,\x')$ given by

\begin{equation}
    c(\x,\x') \equiv -\frac{\delta\beta\Omega}{\delta\rhox\delta\rho(\x')}\bigg|_{\rhob}.
\end{equation}

We remark that the above definition also contains the ideal gas contribution 
to $c(\x,\x')$. For a homogeneous fluid, the direct correlation function is 
isotropic, so $c(\x,\x') \equiv c(\cos\theta)$, where $\theta = \arccos{(\x\cdot\x')}$, 
and we can write it as a spherical harmonics series

\begin{equation}
\begin{aligned}
    c(\cos\theta) = \sum_{\ell} c_{\ell} \Y_{\ell,0}(\cos\theta)\\
    c_{\ell} = \int_{S^2} \dx\,c(\cos\theta)\Y_{\ell,0}(\cos\theta).
\end{aligned}
\end{equation}

We can then use the spherical convolution theorem to rewrite the condition \eqref{stability} as

\begin{equation}
    \sum_{\ell,m} \hcoeff c_\ell (\delta\rho)_{\ell,m}^2 < 0
\end{equation}

which, thanks to the fact that fluctuations $\delta\rho(\x)$ can be chosen arbitrarily, becomes
$\hcoeff c_{\ell} < 0$ for any $\ell$. Finally we can compute these coefficients explicitly to obtain

\begin{equation}\label{corr}
    \hcoeff c_{\ell} = -\frac{1+\etab}{\rhob(1-\etab)^3}-\beta\hcoeff w_{\ell} < 0.
\end{equation}

If the non singular part of the interaction potential has any negative harmonic 
coefficients $w_{\ell}$, one can find some thermodynamic state for which the condition 
is not satisfied at least for some harmonic degree $\ell$ \cite{Likos_2001, Franzini_2018}. 
Then there will be a negative minimum of the coefficients $w_{\ell}$ at some harmonic order 
$\ell^*$ which defines the largest region in the phase diagram in which the homogeneous solution 
is unstable. This region will then necessarily be populated by some inhomogeneous phases which
are stable solutions of eq. \eqref{EulerLagrange}, so that the existence of a negative minimum 
$w_{\ell^*}$ in the harmonic coefficients of the potential becomes the criterion for microphase separation 
to be possible. The boundary of this instability region is defined by

\begin{equation}\label{eqlambda}
    \frac{1+\etab}{\rhob(1-\etab)^3} = -\beta
{\sqrt{\frac{4\pi}{2\ell^*}+1}} w_{\ell^*},
\end{equation}
and it takes the name of $\lambda$-line \cite{Archer_2004}. Notice that while stability is a necessary 
condition for a phase to be the equilibrium state of the system, it is not a sufficient 
condition: this means that, while the homogeneous fluid is stable at temperatures above that of the $\lambda$-line for the $\rhob$ at hand,
it is not necessarily the equilibrium solution. 

As shown in the inset of figure \ref{competing_pot}, $w_\ell$ does indeed display a 
negative minimum at $\ell^* = 6$ for $\sigma/R = 0.1$ and at $\ell^*= 3$ for $\sigma/R = 0.2$, 
which means the phase diagram of the SALR system contains inhomogeneous phases.

\subsection{Monte Carlo simulations}

In order to test the reliability of our mean field approximation, we will use
Monte Carlo (MC) simulations of the SALR model in the canonical ensemble, that is at fixed number
of particles $N$, surface area $A$, and temperature $T$, following the same framework we
used for the GEM4 fluid \cite{Franzini_2018}.

The moves of the MC are rotations of a particle around a random axis orthogonal to its position
vector. More specifically we consider a particles of coordinates $\x = [\cos\phi\sin\theta, \sin\phi\sin\theta,\cos\theta]$
and use its position as the north pole of a new coordinate system, so that its new coordinates
are $\x_0 = [0,0,1]$. Then, we randomly choose a new value $\cos\theta'$ for its third component
$z$ in the interval $[1,1-\delta]$ (setting the maximum stride of the particle to $\delta=0.4 R$),
and choose a direction for the move $\phi' \in [0,2 \pi]$. The new position in the new
coordinate system is given by $\x_0' = [\cos\phi'\sin\theta',\sin\phi'\sin\theta',\cos\theta']$.
Finally we return to the original coordinate system by applying the rotation that satisfies
$\hat{R}\x_{0} = \x$, so that we obtain the new coordinates as $\x' = \hat{R}\x_{0}'$.

Contrary to the GEM4 model and other soft-core fluids, 
the presence of a hard-core in SALR model presents the additional challenge of avoiding 
metastable configurations in which particles can easily get stuck: obtaining a faithful 
representation of the inhomogeneous phases through simulations at high density can be difficult 
and goes beyond the scope of this paper. Hence our simulations will only focus on the homogeneous fluid 
in the low density regime.
This requires using a low number of particles, so we only sample the interval $N\in[2,150]$.

Each sampled trajectory contains $500\, 000$ steps, but the first $150\, 000$
steps are discarded when computing the correlation functions, in order to allow the system
to thermalize.

\section{RESULTS}

\subsection{Correlation functions and the  $\lambda$-line}

\begin{figure}[t]
\centering
  \includegraphics[width=0.5\textwidth]{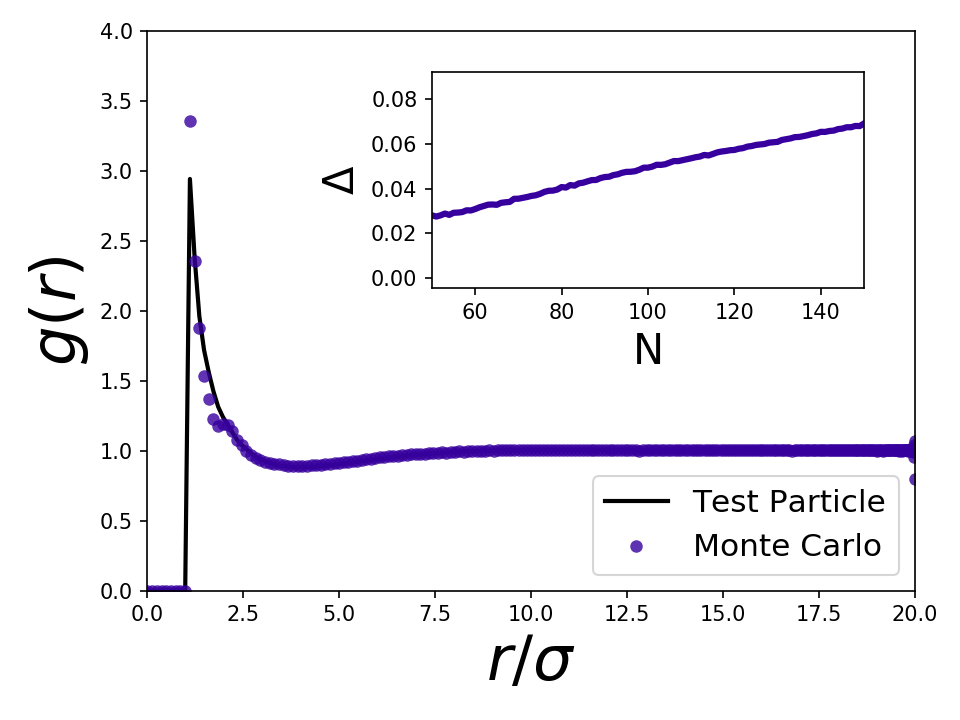}
  \caption{Comparison of the radial distribution function $g(r)$ computed using the 
  test-particle route and Monte Carlo simulations at $T^*=1$, $\sigma/R = 0.1$ and $N=75$ ($\rho R^2 = 5.96$). 
  The inset shows the mean square error $\Delta=\sqrt{\sum ( g^{(TP)}(r_i)- g^{(MC)}(r_i))^2/K}$, with
  $K=256$, between theory and simulation for different values of $N$.}
  \label{gdr}
\end{figure}

Before we consider the full phase diagram, we can characterize the homogeneous fluid. As discussed 
in the previous section, in this regime correlation functions are isotropic and can be computed 
directly from the functional. Once one has computed the direct correlation function $c(\x,\x')$, 
the total correlation function $h(\x,\x')$ can be obtained by solving the Ornstein-Zernike 
equation \cite{Temperley_1977, Tarjus_2011} for the spherical harmonics coefficients

\begin{equation}\label{hcorr}
    h_{\ell} = \frac{c_{\ell}}{1 - \rho \hcoeff c_{\ell}},
\end{equation}
from which one can then compute both the radial distribution function $g(r) = 1 + h(r)$ and the structure factor

\begin{equation}\label{Scorr}
    \hcoeff S_{\ell} = 1 + \rho \hcoeff h_{\ell}.
\end{equation}

However this is not the most accurate path to correlation functions, because the 
LDA reference functional for hard disks gives a direct correlation function proportional
to $\delta(\x,\x')$, so that in \eqref{corr} the hard disk contribution has no dependence on
the harmonic order $\ell$. An alternative route to correlations
is using the Percus test-particle method \cite{Archer_2007}: 
using the functional defined in \eqref{functional}, we consider the action of a 
single particle fixed at the north pole of the sphere on the rest of the fluid, 
that is, we set $\phi(\x) = v(\x)$. Then the equilibrium distribution for states above of the $\lambda$-line
is proportional to the radial distribution function $g(r) = \rhox/\rho$, and from this definition one 
can easily obtain the structure factor $S_{\ell}$ using the previous relations.

\begin{figure}[t]
\centering
  \includegraphics[width=0.5\textwidth]{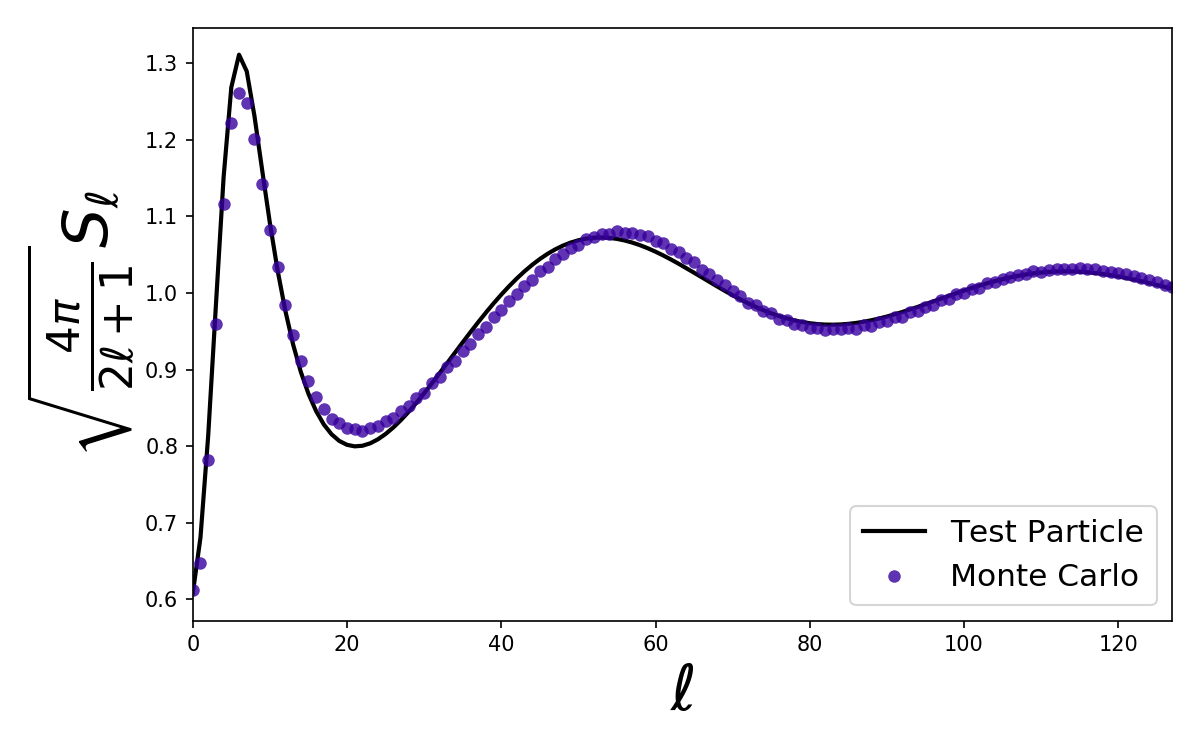}
  \caption{Comparison of the structure factors $S_{\ell}$ computed using the test particle approximation 
  and Monte Carlo simulations at $T^*=1$, $\sigma/R=0.1$ and $N=75$ $\rho R^2 = 5.96$).}
  \label{struct}
\end{figure}

In figure \ref{gdr} we show an example of the radial distribution 
function obtained at $T^*=1$, $\sigma/R=0.1$ and $N=75$, comparing theory and simulation.
Theory is able to describe the first shell of neighboring particles (the peak at $r/\sigma^*=1$) 
and the long-range behavior of the correlation function well enough, however it is not 
able to completely capture the complexity of its short-range behavior: in particular it underestimates
the contact value and does not account for the shoulder at $r/\sigma \simeq 2.5$. 

We quantify 
the agreement of theory and simulations by computing the mean square difference 
between the correlation functions computed with the two methods $\Delta=\sqrt{\sum ( g^{(TP)}(r_i)- g^{(MC)}(r_i))^2/K}$, 
where $K = 256$ is the number of the sampled points. We plot $\Delta$ in the inset of 
figure \ref{gdr}: it shows that on average the error is small with respect to the absolute
value of the sampled point, with differences between theory and simulations 
increasing at very low densities and near the $\lambda$-line.

In figure \ref{struct} we compare the structure factors \cite{Franzini_2018, Anze_2019, Anze_2021} 
obtained from theory and simulations for the same set of parameters. We find good agreement between the two 
at small harmonic orders $\ell$, which account for the large scale 
behavior of the system, and we stress especially that both theory and simulation predict 
the same value for the position $\ell^*$ of the maximum ($\ell^*=6$ in the figure, in agreement
with the prediction of equations \eqref{corr},\eqref{hcorr}, and \eqref{Scorr}). 
This is important because the eventual divergence of this peak at larger densities 
signals the instability of the homogeneous fluid with respect to fluctuations of that 
harmonic order, giving a clue about the characteristics of the inhomogeneous phases 
found  under the $\lambda$-line  (in terms of temperatures). One expects the prediction of the theory to become
unreliable \cite{Archer_2008} at short length scales, corresponding to $\ell \gtrsim \tfrac{2\pi R}{\sigma}$, 
but we find good agreement between theory and simulation even beyond this threshold.

\begin{figure}[t]
\centering
  \includegraphics[width=0.5\textwidth]{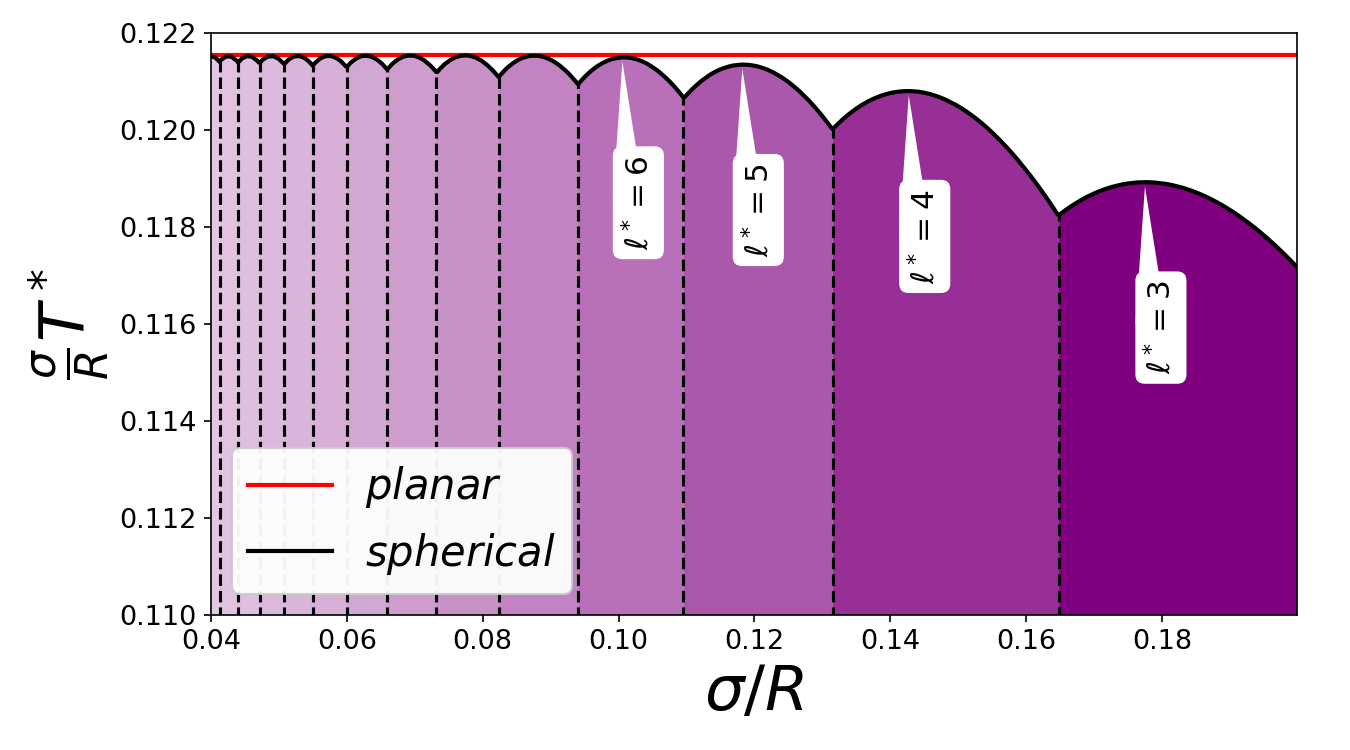}
  \caption{The maximum temperature of $\lambda$-line as a function of the ratio $\sigma/R$. $T^*$ is
  rescaled by a factor $\sigma/R$ in order to isolate the trivial dependency in the planar case and
  better display the difference between planar and spherical systems. The kinks displayed by the $\lambda$-line of the fluid on a spherical surface
  correspond to a change in $\ell^*$: each domain where $\ell^*$ is constant is delimited
  by dashed lines in the figure, and the largest regions are labeled accordingly.}
  \label{lambda}
\end{figure}

The $\lambda$-line itself can be obtained by using equation \eqref{eqlambda}. In figure 
\ref{lambda} we plot the maximum temperature $T^*$ reached by the $\lambda$-line as a function of
the ratio $\sigma/R$, both for the spherical and the planar case. The latter depends trivially
on $\sigma/R$ because of the factor $R$ in the two-Yukawa tail of potential~(\ref{pot}), so we isolate this factor in order to better show the differences between the two
cases.

Similarly to what happens for the GEM4 fluids \cite{Franzini_2018}, $\lambda$-line for the spherical system displays kinks
corresponding to values of $\sigma/R$ where the position of $\ell^*$ shifts (with $\ell^*$ being the position of
the negative minimum of the potential, or of the maximum of the structure factor). This also has consequences on
the structure of the inhomogeneous phases found in each region, although the details of the phase diagram are not
uniquely determined by the position of $\ell^*$.

\begin{figure}[t]
\centering
  \includegraphics[width=0.45\textwidth]{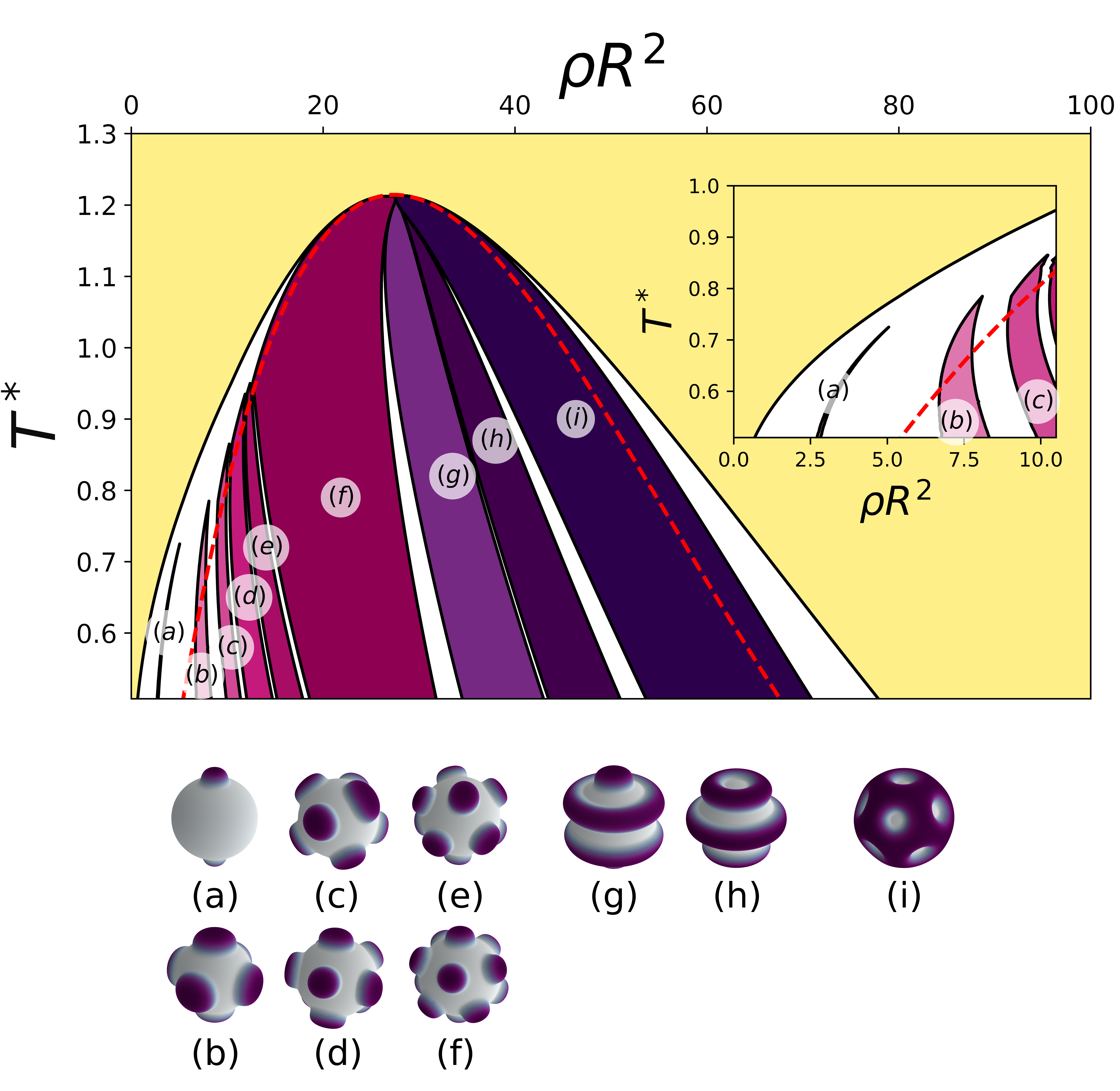}
  \caption{The phase diagram in the $\rho - T^*$ plane for $\sigma/R = 0.1$. Solid lines represent the boundaries 
  between different phases, while the red dashed line is the $\lambda$-line. The white regions 
  are coexistence domains. Aside from the homogeneous fluid (yellow region), $9$ phases appear: 
  (a) $2$ clusters, (b) $6$ clusters, (c) $8$ clusters, (d) $9$ clusters, (e) $10$ clusters, 
  (f) $12$ clusters, (g) $4/3$ stripes, (h) $3/4$ stripes, and (i) $12$ bubbles (we denote 
  stripe patterns by their number of replete/depleted stripes). An example of each phase is 
  provided under the diagram for reference: more dense patches are shown in darker color, 
  moreover a relief is added to the spherical surface to better visualize the density profile.}
  \label{diagram_T}
\end{figure}

\begin{figure}[t]
 \centering
 \includegraphics[width=0.45\textwidth]{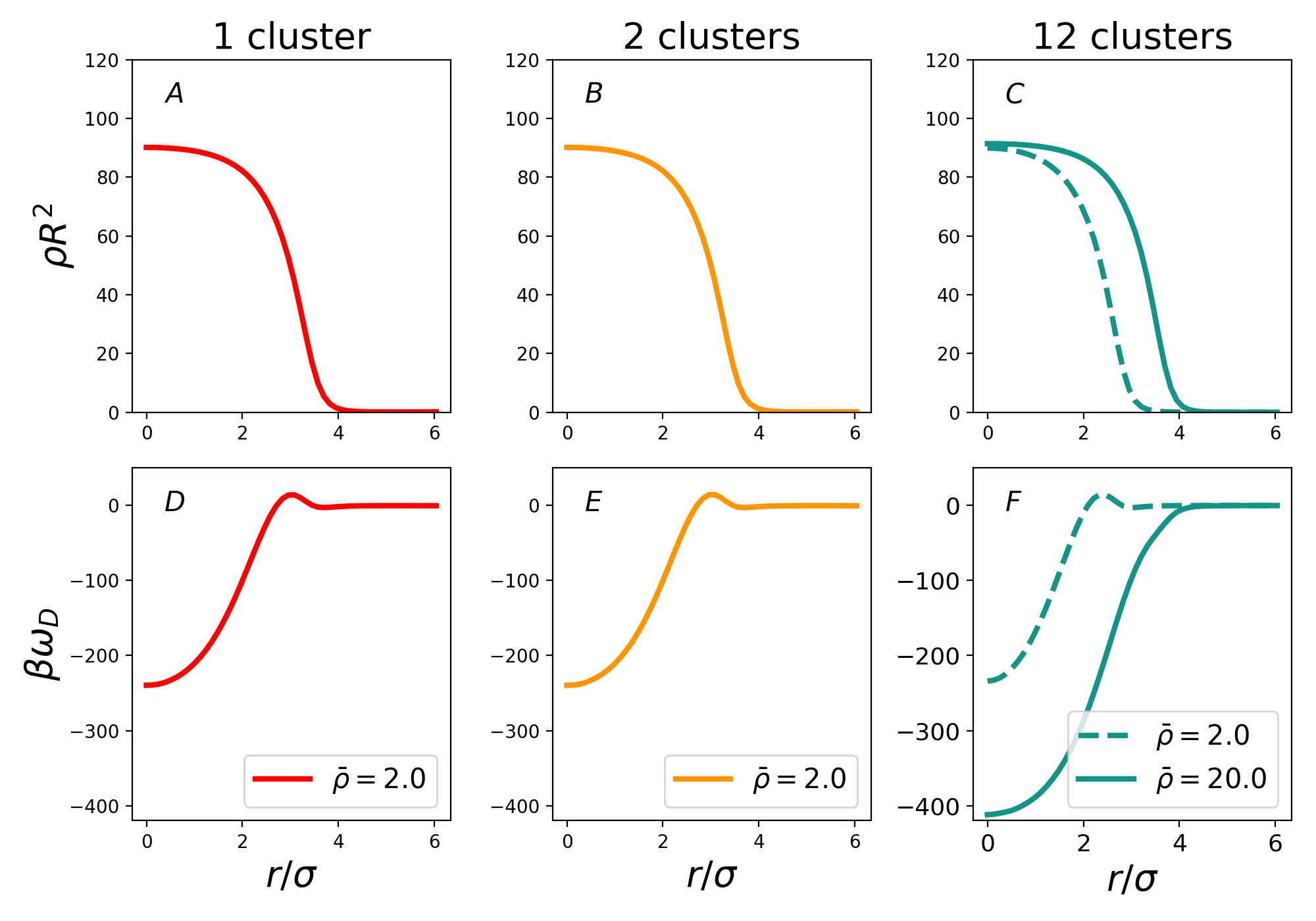}
 \caption{Density profiles for three different cluster crystal phases (1 cluster, 2 clusters, and 12 clusters) at $\sigma/R=0.1$ are shown in the upper panels
 as a function of the distance from the cluster center, while the corresponding local grand potentials profiles are shown in the lower panels. All quantities are
 computed by averaging the values encountered on circumferences centered on the cluster mid-point and having different radii. For the 12-cluster configuration, two different densities are presented in order to illustrate the different behavior at low and high densities. The qualitative difference between the local grand potentials profiles points towards the presence of different mechanisms of cluster stability. }
 \label{pot_bar}
\end{figure}

\begin{figure}[t!]
\centering
  \includegraphics[width=0.45\textwidth]{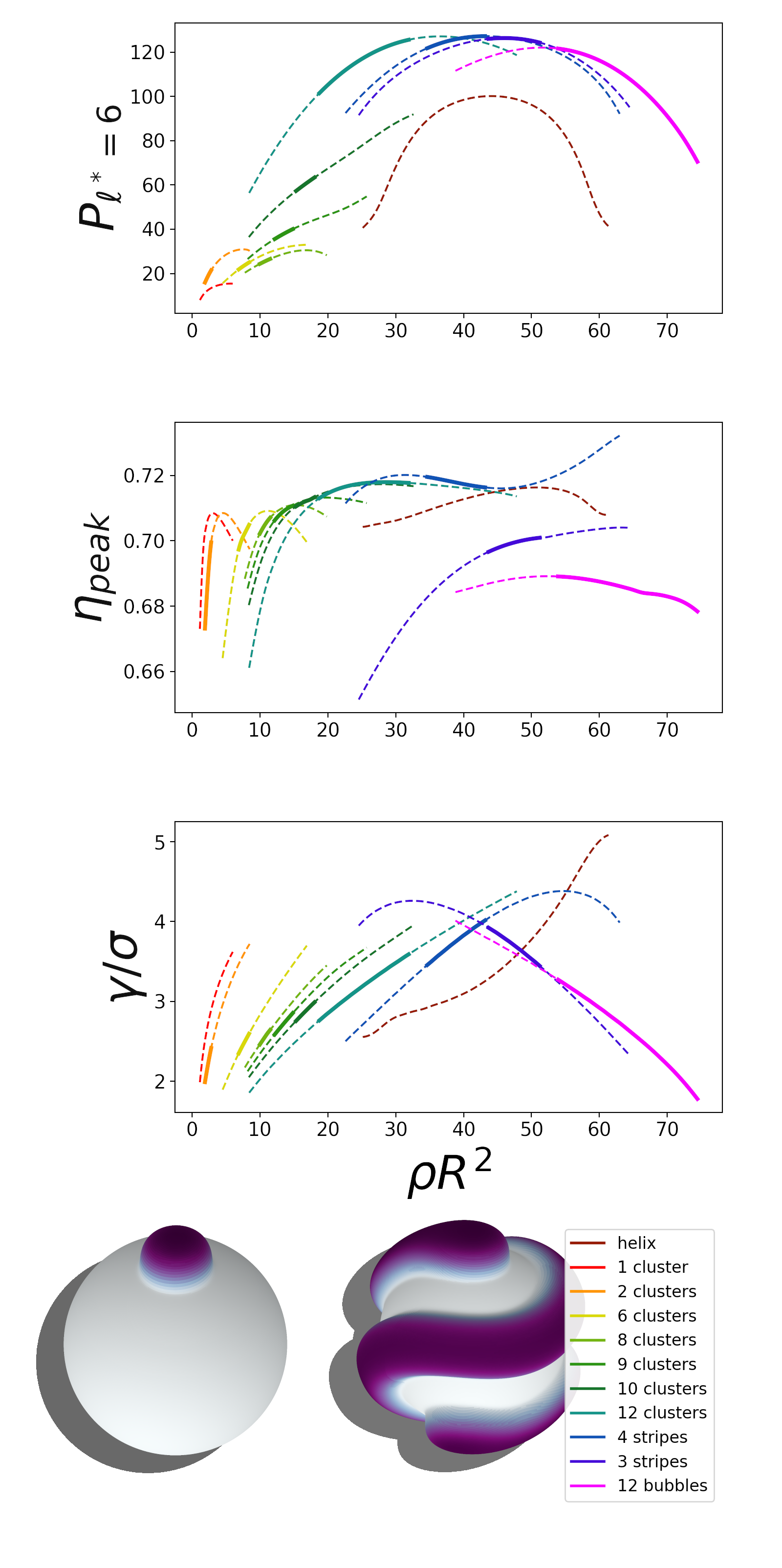}
  \caption{Properties of stable and metastable phases (a single cluster and
  a helix phase) at $\sigma/R=0.1$. The density profiles of the latter are shown below the panels. The top panel shows the total power
  $P_{\ell} =\sqrt{ \sum_{m} \rho^2_{\ell,m} }$ at $\ell=\ell^*=6$. The middle panel shows the maximum
  local packing factor. The bottom panel shows the size of the pertinent mesoscopic structures in units of
  the particles diameter. In each figure, solid lines are used where the phase is the equilibrium one, while
  dashed lines are used where the phase is metastable. Gaps between stable phases correspond to coexistence regions. }
  \label{phase_char}
\end{figure}

\subsection{The phase diagram}

Using the functional developed in the previous sections we can now obtain theoretical predictions 
for the phase diagram of the SALR fluid. We search for the equilibrium density profile for different 
values of the mean density $\rho$, of the temperature $T^*$, and of the ratio  between the size of 
the particles and the radius of the sphere $\sigma/R$.

The boundaries between two phases A and B are obtained by a Maxwell construction, i.e. 
by imposing the conditions $\mu_A = \mu_B$, $\beta\Omega_A = \beta\Omega_B$. The first condition can be
rewritten in terms of the trial densities as $\rhob_A = \rhob_B$. However, since the actual density of
the system $\rho$ is obtained \textit{a posteriori}, we stress that $\rho_A$ is 
not equal to $\rho_B$ at the transition, leading to the appearance of a coexistence region that separates 
the two phases.

We also remark that, as detailed in our previous work \cite{Franzini_2018, Hill_1962}, one 
cannot truly have a phase transition on a sphere, as it is a finite system, and even metastable phases can 
contribute to its thermodynamics. Instead of a discontinuity in the density $\rho$ at the transition 
chemical potential, one expects a smooth crossover of $\rho$, which roughly corresponds to the coexistence regions defined
by the Maxwell construction.

We start by considering the phase diagram in the $\rho - T^*$ plane, at $\sigma/R = 0.1$ corresponding 
to $\ell^*=6$, which we show in figure \ref{diagram_T}.
Fixing a temperature below the critical one and moving from lower to higher densities, the fluid is initially homogeneous. However
as density increases, it undergoes a series of transitions to different cluster crystal phases, 
in which the particles aggregate into clusters arranged on an ordered lattice. Contrary to the bulk 
cases \cite{Pini_2017, Archer_2008}, where only one kind of cluster crystal is found, the finite size of the
sphere leads to the appearance of multiple geometries, characterized by having a different number of clusters ($2$, $6$, $8$, $9$, $10$, or $12$).

As one approaches the center of the diagram, one encounters stripe patterns, in which particles arrange 
in alternating high density (replete) and low density (depleted) stripes around an axis. At $\sigma/R=0.1$ we observe 
two such patterns: one with $4$ replete and $3$ depleted stripes, and another, its reciprocal, having $4$ depleted 
and $3$ replete stripes. We refer to them as the replete and depleted stripe patterns, respectively. Interestingly, 
contrary to what happens in the bulk in two \cite{Archer_2008} and three dimensions \cite{Pini_2017}, the stripe phases do not 
reach the top of the diagram, which is instead occupied by the $12$-cluster crystal phase.

At even higher densities, one finds a bubble phase: particles form a percolating matrix in which depleted spots, 
dubbed bubbles, are arranged in an ordered manner. The $12$-bubble configuration encountered for $\sigma/R=0.1$
corresponds to an inverted $12$-cluster configuration, with bubbles arranged in the same geometry as the clusters.

One can notice a symmetry through the center of the diagram, as phases on either side of the transition between the
two stripe phases are the negative of each other: one stripe pattern is depleted where the other is replete, and vice-versa;
and the same is true for the $12$-cluster phase and the $12$-bubble phase. This, however, is broken by the presence
of the low density cluster phases, which do not have corresponding bubble patterns. An explanation for this behavior will be proposed further on. 

Notwithstanding the differences described above, the overall sequence of cluster, stripe, and bubble phases is similar to that found in the two-dimensional bulk 
\cite{Archer_2008}, and in both cases is driven by the presence of the hard-core part of the potential. While in soft-core fluids such as the GEM4 \cite{Franzini_2018} the density
inside the clusters can grow indefinitely, and clusters actually become more localized as particles are added to the system, in the SALR fluid the hard core prevents particles 
from overlapping, forcing clusters to grow larger and larger. As the density increases, each phase is superseded by another one with higher packing efficiency.   
However, for the case at hand of a system confined on a spherical surface, also finite size effects play a role, and entail a wider geometrical variability of such phases, and the
presence of topological defects.

In the case of cluster and bubble phases, topological disclination defects can be understood as lattice points
having a defective number of nearest neighbors \cite{Tristan_2000}, whereas stripe patterns also contain two disclination defects
located at the poles of the sphere \cite{Mike_1999}. While the variety of geometries encountered in the phase diagram manifest
these topological defects in different ways, all obey the same conservation law which states that \cite{Nieves_2016}

\begin{equation}
    \sum_{k} q_{k} = \frac{1}{2\pi} \int dS\, G = 2,
\end{equation}
where the sum is carried over the topological charges of the defects $q_{k}$, and the integral represents the Euler
characteristic $\xi = 2$ for the $2$-sphere, proportional to the integral of the local gaussian curvature $G$ over the surface.
By computing the topological charge of different kinds of disclinations \cite{Nieves_2016}, one can easily predict the
geometry that will emerge: for instance $12$-cluster crystals must necessarily contain $12$ five-fold disclinations,
because their topological charge is $\tfrac{1}{6}$, while stripe patterns always contain two pole disclinations with
charge $1$.

Another aspect in which the SALR model on a sphere differs from its bulk counterpart is the aforementioned
presence of different cluster crystals for the same value of $\sigma/R$. This also marks a difference with respect to the GEM4
fluid on the sphere considered in a former study \cite{Franzini_2018}: in that case, only one kind of cluster crystal
was observed at each fixed value of $\sigma/R$, independently of density or temperature. More specifically,
our choice of $\gamma_1=\tfrac{5}{6}$ was made to obtain $\ell^*=6$ at $\sigma/R=0.1$: for this
value of $\ell^*$ the phase diagram of the GEM4 fluid only contains configurations with $12$ clusters, whereas 
in the SALR fluid we also observe configurations with fewer clusters, albeit the expected $12$-cluster configuration 
still occupies the largest portion of the region of the phase diagram inhabited by cluster crystals.

The explanation of this more complex behavior could lie in the presence of the short-range attraction, 
which allows particles to aggregate into clusters even at low densities, whereas clustering in purely 
repulsive systems emerges only as a collective behavior at high densities. In fact we even managed to find a 
metastable phase having a single cluster: this configuration never becomes the equilibrium 
distribution, as it is superseded by either the 2-cluster configuration or the homogeneous phase at all densities; 
however, its existence still proves that, contrary to systems with purely repulsive interactions, the attractive 
part of the SALR potential can indeed stabilize isolated structures.

To further illustrate the differences between the low density and high density behavior of the system, we introduce
the local grand potential $\beta \omega_D$, which is the grand potential of a single cell in our grid

\begin{equation}\label{localgpot}
\begin{aligned}
    \beta\omega_D(\x) = \beta f_{id} + \beta f_{HD} + \beta f_{P} - \beta \mu N \\
    \beta f_{id} - \beta\mu N = \frac{\pi^2}{K^2} \rhox\big[ \ln(\rhox/\rhob ) - 1 - \beta\mu_{ex} \big] \sin\theta_{\x} \\
    \beta f_{HD} = \frac{\pi^2}{K^2} \rhox\big[ (1 - \etax)^{-1} + \ln(1 - \etax) - 1 \big] \sin\theta_{\x}   \\
    \beta f_{P} = \frac{\pi^4}{2K^4} \beta \rhox \sin\theta_{\x}  \sum_{\x'} \rho(\x') w(\x,\x') \sin\theta_{\x'}  .
\end{aligned}
\end{equation}

In figure \ref{pot_bar} we plot the density profiles and the local grand potential around the centers of clusters
in different
configurations (1 cluster, 2 clusters, and 12 clusters), and at different densities $\rhob$. 
We recall that $\rhob$ does not generally coincide with 
the actual average density observed in the system, but is an intuitive way of parametrizing the chemical potential. 

Isolated clusters at low densities are all surrounded by an unstable ring, denoted by a positive local grand potential 
at the interface between the cluster itself and the depleted exterior: this is due to a smaller number of neighbors for 
particles at the interface, as well as to the repulsive tail of the potential of particles at the center of the cluster.
However, the presence of a local frustration actually helps the overall structural integrity of the cluster: the number
of frustrated particles is proportional to the diameter $\gamma$ of the cluster, while the surface of the cluster itself
is proportional to $\gamma^2$. Because of this, splitting a cluster into two smaller ones covering the same area leads
to an increase of the number of frustrated particles proportional to $\sqrt{2}$. 

Interestingly, the 12-cluster crystal does not display this characteristic at higher densities: the 
stability of the configuration is ensured by the presence of neighboring clusters pushing against each other, which also 
leads to deeper local grand potential wells. 
Being able to exploit a different clustering mechanism, the SALR fluid displays unexpected cluster phases at low densities, 
but it recovers the same configurational behavior of the GEM4 fluid at higher densities, where collective effects become dominant.
This also offers an explanation for the asymmetry of the phase diagram, i.e. the absence of $2$, $6$, $8$, $9$, and $10$-bubble 
phases: bubble configurations only rely on collective effects for stability, exploiting the long-range repulsions between
particles. This only allows the formation of patterns with the symmetry dictated by $\ell^*$.

\begin{figure*}[h!]
 \centering
 \includegraphics[width=\textwidth]{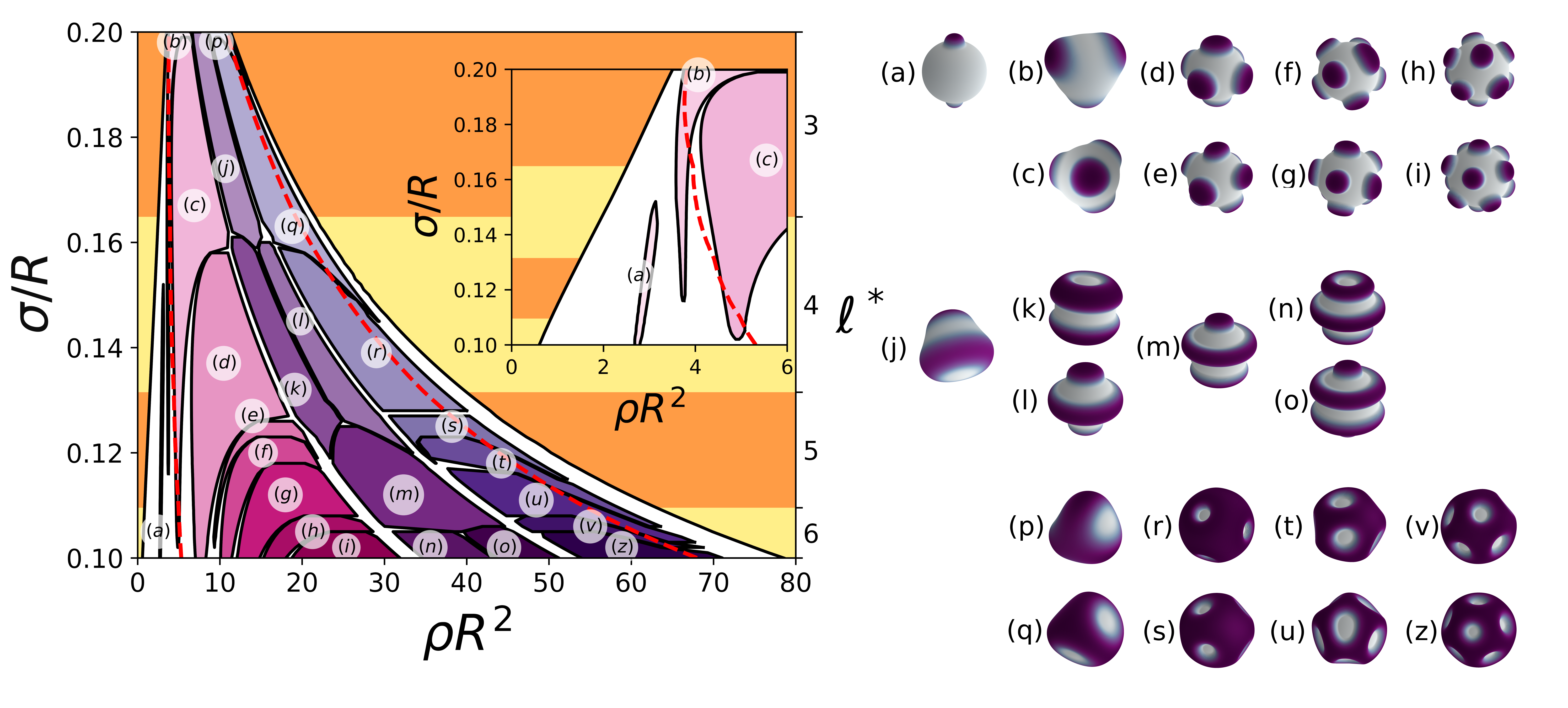}
 \caption{The phase diagram in the $\rho - \sigma/R$ plane at $T^*=0.5$. Solid lines represent the boundaries 
 between different phases, while the red dashed line is the $\lambda$-line. The white regions are 
 coexistence domains. The inset shows a close up of the low density inhomogeneous phases, and the
 right axis enumerates the values taken by $\ell^*$. Aside from the homogeneous fluid, which is shown
 as alternating yellow and orange bands to distinguish different $\ell^*$-domains, $23$ phases appear: (a) $2$ clusters, (b) $3$ clusters, 
 (c) $4$ clusters, (d) $6$ clusters, (e) $7$ clusters, (f) $8$ clusters, (g) $9$ clusters, (h) $10$ 
 clusters, (i) $12$ clusters, (j) $2/2$ stripes, (k) $2/3$ stripes , (l) $3/2$ stripes, (m) $3/3$ stripes, 
 (n) $3/4$ stripes, (o) $4/3$ stripes, (p) $3$ bubbles, (q) $4$ bubbles, (r) $6$ bubbles, (s) $7$ bubbles, 
 (t) $8$ bubbles, (u) $9$ bubbles, (v) $10$ bubbles, and (z) $12$ bubbles (we denote stripe patterns 
 by their number of replete/depleted stripes). An example of each phase is provided next to the 
 diagram for reference: more dense patches are shown in darker color, and a relief is added 
 to the spherical surface to better visualize the density profile. }
 \label{diagram_D}
\end{figure*}

This can be seen in figure \ref{phase_char}, where we analyze some of the quantitative features of the 
inhomogeneous phases, measured at $T^*=0.5$ and $\sigma/R=0.1$. We also compare the properties of the stable 
phases to those of two metastable ones, shown at the bottom of the figure: a single cluster phase and a helix 
phase, which we were able to obtain thanks to the unconstrained minimization algorithm we employed. 
Helix phases have been previously observed in SALR fluids, not only on spherical surfaces \cite{Zarragoicoechea_2009, Pek_2018}, 
but also in 3D systems confined inside pores of various shapes \cite{Pek_2019, Serna_2020}.  
Contrary to what reported for the cases considered there \cite{Zarragoicoechea_2009, Pek_2019}, however, 
the helix phase which we observe is never the equilibrium distribution on the sphere.

In the top panel we compare the total power at harmonic degree $\ell^*=6$, $P_{\ell^*} = \sqrt{\sum_{m}\rho^2_{\ell^*,m} }$ 
of different phases. This quantifies the degree to which the geometry of the observed patterns overlaps with spherical
harmonics of degree $\ell^*$, all characterized by the same overall periodicity.

We observe that, at high density, the range over which a phase is stable roughly 
corresponds to the range of densities at which its $P_{\ell^*}$ is the largest among those of the other phases.
However this is not quite true at lower densities, further emphasizing the presence of two different stability mechanisms between
the two regimes. The helix phase, which is only present at higher densities, has a much lower $P_{\ell^*}$ with respect to the 
other phases and never becomes the equilibrium one. This suggests that phases with a large $P_{\ell^*}$ are better able to exploit 
the characteristic lengthscale introduced by the potential and thus be more stable. 

In the middle panel we show the maximum packing fraction for the various phases. For all configurations $\eta$ 
is much smaller than the close-packing fraction on the plane: this physical behavior is recovered despite the fact 
that $\eta_{CP}\simeq0.9069$ is not a parameter of our density functional theory, in fact according to the 
scaled-particle approximation used here, all values of $\eta$ up to $\eta=1$ would be allowed. Notice that here we refer
to the planar value for the close-packing fraction of hard disks, because its spherical counterpart depends on the
number of particles and is not known for large ensembles \cite{ Kai_2017, caps}.
One can also notice that the densities achieved by the system inside the mesoscopic structures are 
compatible with crystal states \cite{hard3}, although we are not able to probe their internal structure due to the local density approximation. 

Finally, in the bottom panel we show the typical size $\gamma$ of the mesoscopic structures, measured as the half width 
at half maximum of the \textit{high} density patches for the cluster, replete stripe and helix phases, 
and of the \textit{low} density patches for the depleted stripe and bubble phases. Interestingly, none of the 
lower density cluster phases (i.e. all except the $12$-cluster crystal) are stable beyond $\gamma/\sigma=3$, which
roughly corresponds to the position of the maximum of the tail potential, as shown in figure
\ref{competing_pot}. For higher density phases we observe that symmetric patterns (i.e. $12$ clusters and $12$ bubbles, 
or replete stripes and depleted stripes) display a roughly symmetrical behavior of $\gamma$, simply because 
depleted patches become smaller and smaller as particles are added to the system. 

Next we consider the phase diagram in the $\rho - \sigma/R$ plane, at fixed temperature $T^*=0.5$, shown in 
figure \ref{diagram_D}. The homogeneous fluid occupies the low and high density regions, but the middle section
of the diagram is occupied by a plethora of inhomogeneous patterns. We can partition this region into three domains: 
at lower densities we encounter cluster crystals, the middle portion is occupied by stripe patterns, and at high 
densities we find bubble phases. As observed for the phase diagram in the $\rho - T^*$ plane, one can trace an 
approximate symmetry axis passing through the middle of the stripe region, so that phases on opposite sides, excluding
the cluster crystals at low densities, are the negative of each other. In particular one notices that some of the 
stripe phases are their own negative, so that they extend over a larger region on the two sides of the symmetry axis.

We also notice that, as for the previously studied GEM4 fluid \cite{Franzini_2018}, the equilibrium phases are roughly
determined by the value of $\ell^*$. For example, the region of stability of each stripe pattern roughly corresponds
to a single $\ell^*$ domain of figure \ref{diagram_D}. Cluster crystal phases, instead, display
a peculiar behavior, whereby at low densities they extend a tail into regions of lower $\sigma/R$.
This means that cluster phases characterized by a smaller $\ell$ and a larger inter-cluster
distance than the optimal ones for the value of $\sigma/R$ at hand do not disappear abruptly, but 
still survive at low density, in line with the discussion of the case $\sigma/R=0.1$ carried out above.

We conclude our assay of the phase diagram by remarking that the phases we find are qualitatively consistent 
with previous studies of the SALR model through MC simulations \cite{Zarragoicoechea_2009}, albeit some differences 
in the implementations of the SALR potential prevent us from comparing the phase diagram to the phases found 
in that study on a one to one basis. While we are able to find some interesting metastable patterns (such as
the single cluster phase, or the helix phase), we did not observe labyrinth patterns or stable helix phases \cite{Zarragoicoechea_2009,Pek_2019}. 
At least in the case of labyrinth patterns, our inability to locate them may be due to the fact that we 
consider values of $\sigma/R$ much larger than those at which these phases are predicted to occur (i.e. $\sigma/R = 0.01$ \cite{Zarragoicoechea_2009}). 

\section{CONCLUSIONS}

In this work we have studied the phase diagram of a model fluid of hard particles constrained to the surface 
of a sphere interacting via a short-range attractive potential with a long-range repulsive tail, which is known 
to undergo microphase separation in both the 3D and the 2D bulk. We computed the equilibrium configurations 
of a wide set of thermodynamic states, changing the density $\rho$, the temperature $T$, and the ratio 
between the particle diameter and the radius of the sphere $\sigma/R$. To do so we performed a fully unconstrained 
numerical minimization of a simple mean-field functional. The sequence of the equilibrium inhomogeneous phases 
in the phase diagram from lower to higher densities is consistent with previous studies in the bulk \cite{Archer_2008}, 
showing a sequence of clusters, stripes, and bubbles. However, in this case one also needs to consider the additional 
lengthscale of the sphere radius, as well as its topology: these two factors introduce a more complex behavior in the 
phase diagram. One consequence is that even regular patterns include ineliminable topological defects (disclinations) \cite{Nieves_2016}, 
which manifest as structures with a number of neighbors smaller than that expected in the plane for cluster and bubble crystals, and
with pole defects in stripe patterns. The second consequence is 
that by changing the ratio $\sigma/R$ one finds patterns with different numbers of clusters, stripes, or bubbles. 
Nevertheless, similarly to the bulk case, we find that the periodicity of the patterns can still be roughly determined from 
the properties of the non singular part of the interaction potential. Namely, in this case the periodicity is 
related to the harmonic degree $\ell^*$ of the negative minimum of the spherical harmonic expansion of the potential. 
Interestingly, however, we find that this is not the case for the cluster phases at low density, which display density
modulations of lengthscale larger than expected, due to the fact that in these phases clustering is 
achieved via the short-range attraction, rather than through a collective phenomenon mediated by long-range repulsions.

Here we fixed the values of $\gamma_1$ and $\gamma_2$ with respect to the $\sigma/R$ parameter, in order to simplify
the phase diagram. However, it is not necessary to do so: one could as well explore the configurations obtained by
varying these two parameters independently. If this is done by preserving the overall qualitative structure of the
potential (i.e. by still requiring short-range attractions and long-range repulsions), we expect that this operation
would only change the value of $\ell^*$, giving birth to structures (clusters, bubbles, and stripes) with different
periodicity. It may also be possible to tune the parameters in order to make some metastable phases (such as the helix one)
the equilibrium ones, at least in some regions of the phase diagram. If one relaxes the requirements about the specific
shape of the potential, one can also obtain hard-core particles interacting via purely repulsive potentials, which can still 
induce the formation of inhomogeneous patterns \cite{Glaser_2007,Pauschenwein1,Pauschenwein2,Fornleitner_2008,Fornleitner_2010}. 

It is important to note that our theory, while being useful to obtain a qualitative picture of the phase diagram, is still
quite crude with respect to more advanced ones (such as fundamental measure theory), so the results should not be expected
to be quantitatively accurate.

A complementary study to the one presented here could investigate the dynamical aspects of microphase separation 
in the same model fluid, using classical dynamic density functional theory \cite{Tarazona_1999,Archer_Evans_2004} to analyze the way in which the system 
reaches equilibrium and the transitions between different phases. This could lead to interesting results, as 
the ordered equilibrium configurations we found in the phase diagram were in close competition with a multitude 
of metastable phases, such as the helix phase shown in figure \ref{phase_char}, where the system could easily 
become stuck indefinitely for kinetic reasons, as simulations seem to suggest. Another interesting dynamical aspect
that may be studied is particle diffusion between different clusters.

Other future venues of investigation could explicitly include the interaction between the substrate and the particles 
embedded in it, as done for a similar model in a former study \cite{Lavrentovich_2016}. In this picture the short-range attraction could arise as an effective interaction mediated by the curvature induced in the substrate by the 
presence of an embedded particle: one could then study not only microphase separation, but also the resulting shape 
of the substrate at equilibrium.

\section*{Conflicts of interest}

There are no conflicts to declare.

\section*{Acknowledgements}
One of the authors (S. F.) wishes to thank \textit{Laboratorio di Calcolo e Multimedia} (LCM) for providing machine time on their cluster. One of the authors (D. P.) 
acknowledges the financial support by Universit\`a degli Studi di Milano, project No. PSR2019\_DIP\_008-Linea~2.



\balance


\bibliography{rsc} 
\bibliographystyle{rsc} 

\end{document}